\documentclass[prl,showpacs,showkeys,notitlepage,12pt]{revtex4-1}
\usepackage{amsfonts,bbold,amsmath,amssymb,graphicx,epstopdf,verbatim,dsfont,color}
\usepackage[english]{babel}

\newcommand{\be}{\begin{equation}}
\newcommand{\ee}{\end{equation}}
\newcommand{\beq}{\begin{eqnarray}}
\newcommand{\eeq}{\end{eqnarray}}

\begin{document}

\title{Chaos and thermalization in small quantum systems\footnote{This is an unedited version of a perspective on~\cite{Kaufman}; published in Science Vol. 353, Issue 6301, pp. 752-753.}}
\author{Anatoli Polkovnikov$^\ast$ and Dries Sels$^{\ast,\dagger}$}
\affiliation{$^\ast$Department of Physics, Boston University, Boston, MA 02215, USA \\
$^\dagger$Theory of quantum and complex systems, Universiteit Antwerpen, Universitetisplein 1, B-2610 Antwerpen, Belgium}

\maketitle

\date{\today}

{\bf Chaos and ergodicity are the cornerstones of statistical physics and thermodynamics. While classically even small systems like a particle in a two-dimensional cavity, can exhibit chaotic behavior and thereby relax to a microcanonical ensemble, quantum systems formally can not. Recent theoretical breakthroughs and, in particular, the eigenstate thermalization hypothesis (ETH) however indicate that quantum systems can also thermalize. In fact ETH provided us with a framework connecting microscopic models and macroscopic phenomena, based on the notion of highly entangled quantum states.
Such thermalization was beautifully demonstrated experimentally by A. Kaufman et. al.~\cite{Kaufman} who studied relaxation dynamics of a small lattice system of interacting bosonic particles. By directly measuring the entanglement entropy of subsystems, as well as other observables, they showed that after the initial transient time the system locally relaxes to a thermal ensemble while globally maintaining a zero-entropy pure state.}

The laws of thermodynamics are probably the most fundamental in nature as they do not rely on any specific microscopic theory. In particular, the second law postulates that any isolated system will ultimately reach an equilibrium state characterized by the maximum entropy under given macroscopic constraints like total energy, number of particles, and volume. Apart from these constraints all memory of the initial state is lost. This law is intrinsically irreversible as, once a higher entropy state is reached, there is no way to go back as long as the system remains thermally isolated.  At the same time microscopic laws of nature are reversible. This seeming inconsistency was a topic of a big controversy lasting over a century. In classical systems it was partially resolved through chaotic motion happening in generic nonlinear systems, which implies that after a transient time any initial configuration reaches a typical state, on average occupying available phase space points with equal probabilities. However, there are still many open questions like what is this transient time? which configurations are typical? and so on.

Quantum mechanically the situation looks even more confusing as the Schr\"odinger equation, which governs time evolution of the system, is linear and conserves the probabilities of occupying stationary (eigen-) states~\cite{Kaufman,Rigol}. Therefore the density matrix can not possibly relax to any statistical ensemble, always retaining a lot of information about the exact initial state, even when allowing for time-averaging. This simple observation puzzled many generations of physicists. The first person who suggested a possible way out was  one of the pioneers of quantum mechanics J. von Neumann. He formulated ideas of typicality~\cite{Goldstein,Eisert}  realizing that quantum complexity should be hidden in the fact that in macroscopic systems there are exponentially many quantum states. For example, the number of states describing a very small $7\times 7\times 7$ system of spins  is $N=2^{343}$ far exceeding the number of particles in the universe. The physical information one can possibly extract from various observables is far less than what can be encoded in these many states. Consequently, most of the states have to be physically indistinguishable.


This observation, together with random matrix theory, developed and later applied to quantum chaotic systems by E. Wigner, F. Dyson, O. Bohigas, M. Berry and others~\cite{Izrailev}, laid the foundations for the solution. But it was not until the 90's that the role of quantum chaos in emergence of statistical mechanics was finally understood by J. Deutsch and M. Srednicki~\cite{Deutsch,Srednicki1} who formulated a powerful conjecture known as the eigenstate thermalization hypothesis (ETH). In simple words ETH states that a single quantum eigenstate is equivalent to a microcanonical ensemble, in the sense that they make identical predictions about physical observables. These ideas largely remained unnoticed and were initially met with a great deal of skepticism until they were confirmed numerically by M. Rigol et. al.~\cite{Rigol} and in many subsequent works.
The ETH elucidates the connections between microscopic laws and macroscopic phenomena allowing one to make precise and verifiable predictions in chaotic systems~\cite{Luca} . Yet ETH is only a conjecture and there is no known rigorous way to derive it from first principles.

\begin{figure}
  \centering
  \includegraphics[width=\linewidth]{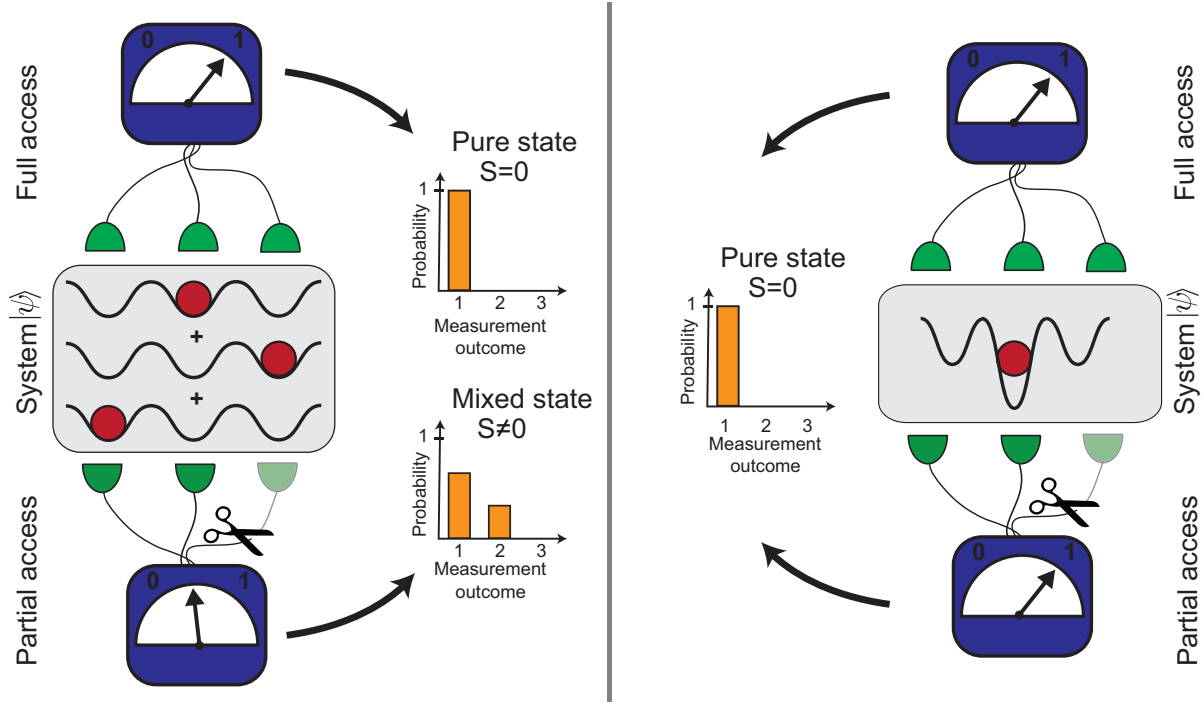}
  \caption{An eigenstate of a particle in a translationally invariant system is a coherent superposition of three localized orbitals (left panel). An observer having access to the whole system can come up with a deterministic measurement
giving reproducible identical results in each experimental realization. Conversely an observer, who can access only sites one and two can only see a statistical mixture of a particle in the state $(|01\rangle+|10\rangle)/\sqrt{2}$ with the probability $2/3$ and the state with no particles $|00\rangle$ with the probability $1/3$ and can not come up with any deterministic measurement. According to ETH, in large chaotic systems the reduced density matrices of stationary states describing small subsystems are maximally mixed exactly like in a thermal ensemble. The level of mixing is encoded in entanglement Renyi entropy $S$ measured in Ref.~\cite{Kaufman}. In disordered systems stationary states are localized (right panel). An observer having access to a subsystem can still identify pure states with low entanglement. Many-body localization phenomenon, recently observed in Refs.~\cite{Schreiber, Choi} implies that surprisingly such non-thermal states can be robust against interactions preventing thermalization even in macroscopic systems. 
}
  \label{fig:main}
\end{figure}

One of the most striking implications of ETH is that thermalization can occur even in relatively small systems. It is only sufficient to have a large Hilbert space, which scales exponentially with the system size. Thus in the system experimentally realized by A. Kaufman et. al.~\cite{Kaufman} consisting of only six bosons confined to six lattice sites the Hilbert space was already 462-dimensional. This allowed the authors to directly verify ETH predictions experimentally for the first time. Specifically they prepared two copies of the same system, with exactly one boson on every site. After a quantum quench, which suddenly allowed particles to hop, correlations grew and the system quickly entangled. By performing a many-body interference experiment on the two copies, suggested in~\cite{Daley} and tested experimentally by the same group~\cite{Islam}, they were able to measure the entanglement entropy of different subsystems as well as the entropy of the full state (see Fig.~\ref{fig:main}). In doing so they beautifully demonstrated that, while the system as a whole remained pure, small subsystems become mixed after a short transient time. In fact, the reduced density matrix of one- and two-site subsystems became indistinguishable from a thermal ensemble.  They also verified this equivalence by direct observation of the particle occupation distribution and comparing it with the equilibrium predictions. On passing, we note that in another recent experiment in even a smaller system of three superconducting qubits~\cite{Neill} a complimentary statement was verified that the full time averaged density matrix becomes thermal in chaotic regimes. This is another direct consequence of ETH~\cite{Luca}.

Not only does ETH validate the use of statistical mechanics, there are also many, potentially important, implications of these ideas to future science and technology. Understanding the microscopic structure of complex systems can give us necessary tools and intuition to get insights into designing systems with a similar or better performance than those found in Nature, e.g. than biological systems, which often operate with amazing efficiency in far from ideal conditions. Comprehending the conditions leading to the breakdown of ETH could be important for developing new technologies not suffering from usual thermodynamic limitations. Remarkably, what first appeared to be an issue of controversy in quantum mechanics ultimately allowed for finding an elegant solution to the problem of thermalization. It is exactly the existence of individual highly entangled eigenstates that allowed one to drop somewhat ambiguous coarse-graining required in standard classical arguments.  Interestingly ETH can be applied to systems near the classical limit, providing a simple mathematical framework to understand unanswered questions in classical chaotic systems.

\acknowledgements
A.P. was supported by NSF DMR-1506340, ARO W911NF1410540 and AFOSR FA9550-16-1-0334. D.S acknowledges support of the FWO as post-doctoral fellow of the Research Foundation - Flanders.

\end{document}